\documentclass[prd,tightenlines,nofootinbib,superscriptaddress]{revtex4-1}

\usepackage{amsmath}
\usepackage{amsfonts}
\usepackage{amssymb}
\usepackage{dsfont}
\usepackage{mathrsfs}
\usepackage{bm,slashed}
\usepackage{graphicx}
\usepackage{color}
\usepackage{hyperref}
\usepackage{MnSymbol}
\hypersetup{
    colorlinks=true, 
    linktoc=page,    
    linkcolor=blue,  
    urlcolor=black
}

\usepackage[ngerman]{babel}
\usepackage{stackengine}
\usepackage{scalerel}

\DeclareFontFamily{U}{rsfs}{}         
\DeclareFontShape{U}{rsfs}{m}{n}{<5> rsfs5 <6><7> rsfs7          %
  <8><9><10><10.95><12><14.4><17.28><20.74><24.88> rsfs10}{}     %
\DeclareMathAlphabet{\mathfs}{U}{rsfs}{m}{n}                     %
\newcommand{\be}{\nopagebreak[3]\begin{equation}}
\newcommand{\ee}{\end{equation}}

\newcommand{\bee}{\nopagebreak[3]\begin{equation*}}
\newcommand{\eee}{\end{equation*}}

\newcommand{\ba}{\nopagebreak[3]\begin{eqnarray}}
\newcommand{\ea}{\end{eqnarray}}

\newcommand{\baa}{\nopagebreak[3]\begin{eqnarray*}}
\newcommand{\eaa}{\end{eqnarray*}}

\newcommand{\la}{\label}
\newcommand{\n}{\nonumber}

\newcommand\myeq[1]{\stackrel{{#1}}{=}}

\def\pa{\partial}
\def\rd{\mathrm{d}}
\newcommand{\va}{\scriptscriptstyle}

\def\be{\begin{equation}}
\def\ee{\end{equation}}
\def\ba{\begin{eqnarray}}
\def\ea{\end{eqnarray}}

\def\label{\langle}

\def\SU{\text{SU}}
\def\ISU{\text{ISU}}

\def\su{\mathfrak{su}}

\newcommand{\e}{\mathrm{e}}
\newcommand{\rt}{\mathrm{t}}
\newcommand{\m}{\mathrm{m}}

\def\cB{{\cal B}}
\def\vphi{\varphi}

\begin{document}

\title{Kinematical Gravitational Charge Algebra}
\date{\today}

\author{{\bf Laurent Freidel}}\email{lfreidel@perimeterinstitute.ca}
\affiliation{Perimeter Institute for Theoretical Physics, 31 Caroline St. N, N2L 2Y5, Waterloo ON, Canada}

\author{{\bf Etera R. Livine}}\email{etera.livine@ens-lyon.fr}
\affiliation{Perimeter Institute for Theoretical Physics, 31 Caroline St. N, N2L 2Y5, Waterloo ON, Canada}
\affiliation{Universit\'e de Lyon, ENS de Lyon, Universit\'e Claude Bernard, CNRS, LP ENSL, 69007 Lyon, France}

\author{{\bf Daniele Pranzetti}}\email{dpranzetti@perimeterinstitute.ca}
\affiliation{Perimeter Institute for Theoretical Physics, 31 Caroline St. N, N2L 2Y5, Waterloo ON, Canada}

\begin{abstract}

When formulated in terms of connection and co-frames, and in the time gauge, the phase space of general relativity consists of a pair of conjugate fields: the flux 2-form and the Ashtekar connection. On this phase-space, one has to impose the Gauss constraints, the vector, and scalar Hamiltonian constraints. These are respectively generating local SU$(2)$ gauge transformations, spatial diffeomorphisms, and time diffeomorphisms. We write the Gauss and space diffeomorphism constraints as conservation laws for a set of boundary charges, representing spin and momenta, respectively. We prove that these kinematical charges generate a local Poincar\'e ISU$(2)$ symmetry algebra. This gives strong support to the recent proposal of Poincar\'e charge networks as a new realm for discretized general relativity \cite{Freidel:2019ees}.

\end{abstract}

\maketitle

\section{Introduction}

General relativity is a fully constrained theory of spacetime geometry, which means that its whole physical content is encoded in its gauge transformations and the resulting symmetries and conserved charges induced on the boundaries of space-time. The goal of the present work is to revisit the Hamiltonian analysis of general relativity in its first order formulation \`a la Cartan in terms of coframe and connection, with a special focus on boundary terms and symmetries.

This analysis is based on a new perspective in the way one handles differentiability of gauge generators in the presence of boundaries. In this approach one allows for the presence of boundary degrees of freedom, the edge modes, which are acting upon by the boundary  symmetry charges. There is growing understanding that edge modes must play a key role in quantum gravity. They are central in the quantization of 2d Yang-Mills and 2d gravity \cite{Blommaert:2018oue,Blommaert:2018iqz, Donnelly:2019zde}, They play a key role in 
 3d quantum gravity \cite{Afshar:2016kjj,Geiller:2017whh,Freidel:2018pbr,Goeller:2019zpz}.
 They are essential to   the understanding of boundary dynamics and to the construction of  defects operators in Chern-Simons theory \cite{Elitzur:1989nr,Banados:1994tn,Geiller:2017xad,Wong:2017pdm}.
 In four and higher space-time dimensions, edge modes  have been argued to  be essential to understanding black hole entropy \cite{Bekenstein:1994bc,Carlip:1996yb,Carlip:1998wz,Haco:2018ske}. 
 Finally they are now understood to be a key ingredient in the holographic nature of gravity \cite{Regge:1974zd,Freidel:2013jfa,Donnelly:2016auv,Balachandran:2018lcf}
 and they provide a new understanding in the quantization of geometric observables\cite{Freidel:2015gpa,Freidel:2016bxd,Freidel:2019ees}.

In  \cite{Freidel:2019ees},  relying on the assumption that curvature on the boundary is distributional  (i.e. it vanishes on the boundary, except at the location of a given set of points or punctures), we exploited the local holographic nature of gravity to
put forward an extension of the kinematical quantum degrees of freedom given by the spin networks in Loop Quantum Gravity. 
The analysis of  \cite{Freidel:2019ees},  motivated by a realization of quantum gravity as dynamical networks of quantum edge modes,
led to a novel construction of tubular networks dressed by representations of the Euclidean  $\mathfrak{iso}(3)$ algebra generated by the fluxes and momentum operators (plus the additional higher mode quantum numbers). This is done by gluing together 3D regions bounded by 2D surfaces with punctures where the edge modes live and which are presumed to carry all the dynamical information of the 3D bulk geometry. The most natural reduction or coarse-graining which traces out the higher modes of these tubular networks leads to these generalized spin networks, dubbed {\it Poincar\'e networks}, which then carry a representation of the SU(2) charges from the Gauss constraint and also of the momentum charges associated with the diffeomorphism constraint.

In the present paper, we review the classical setup and provide further details, both at the conceptual and the technical level, of the novel framework developed in  \cite{Freidel:2019ees}.
The key idea we develop is two-fold. First 
we show that the constraints on the phase space of general
relativity can be written as conservation laws of local charges.
These charges can be integrated along a 2-dimensional surface after smearing with local symmetry parameters.
Then we also show that  the conserved charges, properly supplemented with boundary terms, do generate the correct gauge transformations.  
These boundary terms involve edge modes propagating on the boundary and carrying a boundary symplectic form. 
The edge modes  are would-be-gauge degrees of freedom induced on the boundary, shadows of the bulk gauge transformations,  which ensure that the conserved charges are properly differentiable with respect to field variations, both in the bulk and on the boundary.

In our analysis, we assume the time gauge, i.e. we fix the time direction in the internal space to the 4-vector $n^I=(1,0,0,0)$, which partially gauge-fixes the local Lorentz transformations down to local $\SU(2)$ transformations. The phase space of general relativity then consists of the canonical pair of conjugate fields, the Ashtekar-Barbero $\SU(2)$ connection $A$ and the co-triad or flux 2-form $\Sigma=e\wedge e$, supplemented with first class constraints generating local $\SU(2)$ gauge transformations and space-time diffeomorphisms.
We focus on the kinematics of the theory, analyzing the $\SU(2)$ gauge transformations and space diffeomorphisms while postponing  studying the fate of time diffeomorphisms and the time evolution of the geometry to future work.
In the first part of this work, we show that the constraints generating the $\SU(2)$ gauge transformations and space diffeomorphisms can be recast  in terms of conservation of boundary charges satisfying a Poincar\'e algebra $\ISU(2)$. While the $\SU(2)$ sector of Poincar\'e obviously corresponds to the local gauge transformations, the space diffeomorphisms are now written as field dependent translations.

More precisely, we consider a bounded 3D region $\cB$ within the canonical space-like Cauchy hypersurface and call its boundary $S$.
We establish the  validity of the Poincar\'e algebra for the boundary charges $G_{\alpha}$ and $P_{\varphi}$ which are respectively the local generators of  local internal rotations and  translations:
\be
\{G_\alpha,G_\beta\}= G_{(\alpha\times \beta)}\,, \qquad \{G_\alpha,P_\varphi\}= P_{(\alpha\times \varphi)}\,,\qquad \{P_\varphi,P_{\xi}\}=0
\,,
\ee
where  $\alpha,\beta,\varphi,\xi$ are $\su(2)$-valued scalar fields on the hypersurface
and $\alpha\times \beta$ denotes the cross product. 
Now, coming back to the diffeomorphisms, we can map an arbitrary vector field $\hat{\varphi}=\hat{\varphi}^a\pa_a$ to a $\su(2)$-valued scalar field $\vphi^i=\iota_{\hat{\vphi}}e^i=\hat{\vphi}^ae^i_{a}$ and identify the charge corresponding to the 3D diffeomorphism as a field-valued translational charge:
\be
D_{\hat{\vphi}}=P_{\iota_{\hat{\vphi}}e}.
\ee
This allows us to show that, while the translational charges commute with each other, the covariant diffeomorphism algebra admits an extension proportional to the boundary curvature on $S$:
\be 
\{D_{\hat{\xi}},D_{\hat{\varphi}} \}=D_{[\hat\xi,\hat\varphi]}+G_{\iota_{\hat \xi}\iota_{\hat \varphi}F}.
\ee

This realization of the algebra of conserved charges as a Poincar\'e algebra comes in support of the recent proposal of discretizing general relativity in terms of Poincar\'e charge networks \cite{Freidel:2019ees}.

In the section \ref{sec:bulk} below, we review the bulk phase space for general relativity with the constraints written as conservation laws for electric, magnetic and translation charges. In the following section \ref{sec:boundary}, we introduce the edge modes and correct the conserved charges with the appropriate boundary terms. We show that they generate as expected the $\SU(2)$ gauge transformations and 3D diffeomorphisms. In the final section \ref{sec:algebra}, we compute the algebra of charges and show that they are effectively repackaged as a Poincar\'e algebra.

\section{Bulk phase space and constraints}
\label{sec:bulk}

Given a  space-like Cauchy hypersurface $M$ in a 3+1 decomposition of  spacetime, we consider a bulk  3D  region $\cB\subset M$ with a 2D boundary $S$. The bulk phase space of first order gravity in connection formulation is parametrized by the Ashtekar--Barbero $\SU(2)$ connection and the $\su(2)$-valued flux 2-form, namely 
\be
A^i := \Gamma^i+\gamma K^i\,,\quad \Sigma_i=\frac12 (e\times e)_i\,,
\ee
where we denote the $\SU(2)$ bracket $(f\times g)^i=\epsilon^i\!_{jk}f^j \wedge g^k$ for arbitrary $\su(2)$-valued forms $f$ and $g$, with the indices $i,j,k\in \{1,2,3\}$ labeling the Pauli matrices as a basis of the $\su(2)$ Lie algebra.
Above, $\Gamma^i$ is the 3d spin connection, $e^i$ the normalized\footnote{This means that we have rescaled the coframe field by $e^i \to \frac{e^i}{\sqrt{\kappa \gamma}}$.} 3d-frame field, and $K^i$ the extrinsic curvature one-form:
\be
\rd_\Gamma e^i=0,\qquad  K^i:= \rd_\omega n^i\,,
\ee
with $n^i$ the hypersurface internal normal and $\omega$ the Lorentz connection. In the time gauge we have $e^0=n$.

The Ashtekar-Barbero connection and the flux 2-form form a pair of conjugate fields, so that the bulk pre-symplectic 2-form reads (see \cite{Thiemann:2007zz} for a review of the canonical analysis of general relativity):
\be
\la{Obulk}
\Omega_{\cB} =  \int_{\cB} ( \delta A^{i}\wedge \delta
\Sigma^{\va}_{i} )
\,.
\ee
We focus our attention on the kinematical theory, which means that the phase space is restricted by the action of two sets of kinematical constraints: the Gauss law and the space diffeomorphism constraints. 
We postpone  to future  work the investigation  of the full dynamical theory that takes into account the time diffeomorphism constraint.
These kinematical constraints are usually \cite{Thiemann:2007zz} written in terms of the canonical variables $A_a^i$ and the densitized triad $\tilde{E}^a_i:= \epsilon^{abc}\Sigma^i_{bc}$ as
\be
G_i=\nabla_a \tilde{E}^a_i=0 ,\qquad D_a = F^i_{ab}\tilde{E}^b_i=0.
\ee
It is nevertheless convenient to keep working with differential forms, in which case the Gauss law is written as a conservation law
$G_{i}:=\rd_A \Sigma_i=0$  for the Lie algebra valued two-form $\Sigma^i$ identified as the electric charge aspect.
A central point of our approach \cite{Freidel:2019ees} is to show that the diffeomorphism constraint can similarly be written as a conservation law for a momentum aspect defined as:
\be\la{P0}
P_i:=\rd_A e_i\,.
\ee
To do so, we use an isomorphism between vector fields, and Lie-algebra valued functions
\ba
 \Gamma(T \cB) &\to& C(\su(2))\cr
 \hat{\varphi} &\mapsto & \varphi^i:= \iota_{\hat{\varphi}} e^i \,.
\ea
In the following, we will denote vector fields with a hat $\hat{\varphi}=\hat{\varphi}^a\pa_a$, which should be distinguished from the corresponding scalar functions $\varphi^i := \iota_{\hat\varphi} e^i= \hat \varphi^a e_a^i$.
The diffeomorphism constraint associated with the vector field $\hat\varphi$ is denoted 
$D_{\hat\varphi}$.
We can now witness the ``caterpillar to butterfly'' transformation of the 3D diffeomorphism generator:
\ba
D_{\hat{\varphi}}&=& (\hat\varphi^a F^i_{ab} \tilde{E}^b_{ i})\cr
&=& \iota_{\hat{\varphi}} F^i \wedge \Sigma_i
\cr
&=&
-F^i\wedge \iota_{\hat{\varphi}}\Sigma_i\cr
&=&
F^i\wedge (e\times \iota_{\hat\varphi} e )_i \cr
&=& (F \times e)_i\, \varphi^i \cr
&=&(\rd_A^2 e_i)\,  \varphi^i\cr
&=&(\rd_A P_i)\,  \varphi^i\,,\label{butterfly}
\ea
where we have used the useful relation $(F\times \eta)^i=\rd_A^2 \eta^i$ valid for any $\su(2)$-valued $n$-form $\eta^i$.
This shows that the diffeomorphism constraint follows from a conservation law, namely the conservation of momenta $\rd_A P_i=0$.

This momentum aspect $P_{i}=\rd_A e_i$ actually measures  the torsion of the Ashtekar-Barbero connection $A$. Let us remember that the Ashtekar-Barbero connection $A=\Gamma+\gamma K$ combines the 3D spin-connection $\Gamma$ and the extrinsic curvature $K$. Since the spin-connection is by definition torsionless, $\rd_\Gamma e=0$,  the Ashtekar-Barbero torsion $\rd_A e_i$ simply measures the extrinsic curvature up to a factor given by the Immirzi parameter, $P=\rd_A e=\gamma (K \times e)$.
This provides a direct geometrical interpretation of the momentum aspect 2-form $P$ in terms of the extrinsic curvature of the hypersurface.

\medskip

Putting the Gauss law $\rd_A \Sigma^i=0$ together with the 3D diffeomorphism constraint $\rd_{A} P^i=0$, we get two conservation laws, to which it is natural to add the Bianchi identity $\rd_A F^i=0$ satisfied automatically by the Ashtekar-Barbero connection. This means that the theory at the kinematical level is defined by the following three conservation laws  in the bulk:
\begin{align}
&\rd_A \Sigma^i\simeq0\,,&\text{Electric Gauss law}\la{G-cons},\\
&\rd_A P^i\simeq0\,,&\text{Translation constraint}\la{P-cons},\\
&\rd_A F^i\simeq0\,,&\text{Magnetic Gauss law}\la{F-cons}.
\end{align}
The first two are first class constraints to be imposed on the phase space variables, while the last one is a topological constraint directly implied by the definition of the curvature tensor $F(A)$ and the exterior derivative $\rd$ for the differential calculus.
An interesting  relationship exists between these constraints:
\be
\rd_A \Sigma= P\times e, \qquad \rd_A P= F\times e, \qquad \rd_{A}F =0
\,,
\ee
which hints at an intriguing hierarchy and order between these conservation laws.
The first identity says that any source to the Gauss constraint is a source of angular momenta since $P\times e$ can be understood as the angular momenta density associated with the momenta $P$.
The second equality is more surprising as it suggests that the momenta density, which appears on the RHS of the momentum constraints, is itself an ``angular momenta'' associated with the monopole charge aspect, or curvature  $F$.
Note that the last equations stay sourceless since any source would be a gravitational monopole, which is excluded.
The full meaning of these equations remains to be unraveled. 

%
%
The associated charges that are covariantly conserved are  the electric,  the translational and the magnetic charges\footnote{
The conservation law in the bulk $\rd_{A}\Sigma=0$ implies by an integration by parts that the boundary charge is given by the bulk integral of the covariant variation of the electric gauge parameter:
\be
\rd_{A}\Sigma=0
\quad\Rightarrow\quad
Q^{\e}_\alpha=\int_S \alpha^i \Sigma_i = \int_{B}\rd_{A}\alpha^i\wedge \Sigma_{i}
\,.
\nonumber
\ee
If the electric gauge parameter is held covariantly constant in the bulk, then the boundary charge vanishes $Q^{\e}_{\alpha}=0$. In particular, if the boundary consists in two disjoints parts, e.g. if the 3D bulk is an open cylinder between an initial surface and a final surface, then the initial surface charge and final surface charge are equal. The same holds for the translational and magnetic boundary charges.
}

\begin{align}
&Q^{\e}_\alpha=\int_S \alpha^i \Sigma_i\,,\qquad Q^{\rt}_\varphi=\int_S\varphi^i P_i\,,
\qquad
Q^{\m}_\beta=\int_S\beta^i F_i\,,
\end{align}
where $\alpha^i$ and $\beta^i$ respctively denote the $\SU(2)$ electric and magnetic gauge parameters.
Using the isomorphism between vector fields $\hat{\vphi}=\hat{\vphi}^a\partial_{a}$ and $\su(2)$-valued scalar functions $\vphi^i$, the diffeomorphism charge can then be written as a field dependent translation associated with the vector $\hat\varphi$ 
\be
Q^{\rd}_{\hat\varphi}:= Q^{\rt}_{\imath_{\hat\varphi}e }= \int_S\imath_{\hat\varphi}e^i \rd_A e_i= \frac12 \int_Se^i L_{\hat\varphi} e_i
\,,
\ee  
where $L_{\hat\varphi}:= \rd_A \iota_{\hat\varphi}+\iota_{\hat\varphi}\rd_A $ is the covariant Lie derivative along the vector field $\hat{\vphi}$.

\medskip

As it is well-known,  any first class constraint also plays the role of  canonical generators for an associated gauge transformation. This dual role of the constraints, often phrased by pointing out that each first class constraint kills two degrees of freedom, reflects the fact that initial data differing by an infinitesimal gauge transformation again solve the constraints. In other words, $(A, \Sigma)$ and an infinitesimal variation $(A+\delta A, \Sigma+\delta \Sigma)$ represent the same data but in different gauges, as long as 
 \be\label{transf}
 \delta_\alpha A=\{ A, H_\alpha \}\,,\quad  \delta_\alpha \Sigma=\{ \Sigma, H_\alpha\}\,,
 \ee
where $H_\alpha$ is the Hamiltonian generating the given transformation $\delta_\alpha$, namely
\be\la{Ham}
\delta H_\alpha = \Omega_{\cB}(\delta_\alpha, \delta  ) =I_{\delta_\alpha}\Omega_{\cB} \,,
\ee
where we denote by $I$  contraction in field space, namely $I_\delta \Omega$ represents the contraction  between the (field space) vector field $\delta$ and the (field space) 2-form $\Omega$.

In order for $\delta_\alpha$ to be a gauge transformation one needs 
its hamiltonian generator $H_\alpha$ to vanishes identically on solutions.
On the other hand, the Hamiltonian generator of a symmetry transformation that does not vanish on-shell generates  truly physical canonical transformations that change the system.
In the presence of a boundary, it is well-known that some of the would-be-gauge transformations are in fact symmetries. This becomes especially important when analyzing the edge modes propagating on the boundary, which are the modes conjugated to the boundary symmetry generators.
The Poisson bracket of two Hamiltonian generators is given by:
\be
\{H_\alpha,H_{\beta} \} = \delta_{\beta}   H_{\alpha}= \Omega_{\cB}(  \delta_\alpha, \delta_\beta)\,,
\ee
where $\delta_\alpha, \delta_\beta$ are the Hamiltonian vector fields
generated by the two Hamiltonians. 

As such, the covariant phase space formalism provides efficient tools to identify the gravitational gauge charges and to study their algebra. An important feature of this formalism is the requirement that the Hamiltonian generator  be differentiable with respect to (all) field variations, as demanded by the definition of the Poisson bracket. This requirement has far reaching implications in the presence of boundaries \cite{Regge:1974zd}. To understand this crucial aspect clearly, let us first notice that differentiability of the Hamiltonians implies that some of the constraints need to be integrated by parts. For instance, when the bulk region $\cB$ is bounded by a surface $S$, the generator of electric gauge transformations requires an appropriate boundary term in order to take a differentiable form:
\be\la{Gb}
G_\alpha
=
\int_{\cB} \alpha^i\wedge \rd_A\Sigma_i
-\int_{S}\alpha^i \wedge \Sigma_{i}
=
-\int_{\cB} \rd_A\alpha^i\wedge \Sigma_i
\,.
\ee 
We see  that in the presence of a boundary and in the absence of edge modes\footnote{That is, in the absence of the introduction of explicit new boundary degrees of freedom.}, the gauge transformations are only the ones associated with a gauge parameter vanishing at the boundary,
while symmetries correspond to transformations with non-vanishing boundary parameter.
We could then follow the same strategy also for the translation constraint \eqref{P-cons} and the magnetic Gauss law \eqref{F-cons} and introduce the corresponding bulk generators
\ba
P_\varphi&=&-\int_{\cB} \rd_A \varphi^i\wedge \rd_A e_i\,,\la{Pb}\\
F_{\beta}&=&-\int_{\cB}  \rd_A\beta^i\wedge F_i(A)\,.\la{Fb}
\ea
However, even after this integration by parts, the generators \eqref{Pb}, \eqref{Fb} are still not differentiable, as their variation still yields a boundary contribution.
At this point, to define the generators of translations and magnetic transformations as proper Hamiltonians, one can follow two strategies.
The first option is the most standard: one imposes boundary conditions by demanding that the phase space field variations vanishes on the boundary. 
 In particular, $ \delta e^i\myeq{S}0$  ensures the differentiability of the translation generators.
 While  $\delta A^i\myeq{S}0$ ensures the differentiability of the magnetic gauss generator (here $\myeq{S}$ denotes an equality for forms pulled-back to $S$). However, such substantial restrictions can kill boundary degrees of freedom, which may play an important physical role (see, e.g., \cite{Tong:2016kpv} for the role of edge modes in the description of the quantum Hall effect).
Physically, this means that we consider  physical processes restricted to a charge superselection sector. That is processes that do not change the value of the boundary charges: $\delta Q_\alpha=0$. 
The second strategy aims to allow such processes and more flexible boundary conditions by extending the bulk phase space with boundary edge modes.
This allows for a physical interpretation of the boundary charges, in particular for momentum charges associated with the diffeomorphism constraint, and a symmetric treatment of the Gauss and diffeomorphism constraints.
 We describe in detail this alternative approach in the next section.

\section{Extended phase space}
\label{sec:boundary}

In this section we perform a phase space extension by introducing the electric edge modes. This allows us to define differentiable Hamiltonian generators of  both electric and translational gauge transformations, which we compute  explicitly. 
We postpone to future investigation the study of magnetic gauge transformations associated to the Bianchi identity.

\subsection{Boundary pre-symplectic 2-form}

In order to allow for boundary field variations, we can extend the bulk phase space by adding a boundary term to the pre-symplectic 2-form \eqref{Obulk}. In the present work, we focus on electric gauge transformations and translations, and we postpone the detailed study of magnetic gauge transformations to future investigation. This means that we will allow arbitrary boundary field variations for the triad field $e$ while keeping the Ashtekar-Barbero connection fixed on the boundary.
In this setting, we see from the expression \eqref{Pb} that, in order for the translation generator to become differentiable, we need to parametrize the boundary pre-symplectic 2-form in terms of a boundary  co-frame field $\e^i$, which we distinguish from  the bulk frame $e^i$.
Therefore, if we allow for arbitrary boundary variations $\delta \e^i$, while still keeping the boundary connection fixed $\delta A^i\myeq{S}0$, and thus the boundary curvature fixed, $\delta F^i(A)\myeq{S}0$,  we introduce the extended pre-symplectic 2-form, as shown in the earlier work \cite{Freidel:2015gpa, Freidel:2016bxd}:
\be\la{Oext}
\Omega = \Omega_{\cB}+\Omega_{S}= \int_{\cB} ( \delta A^{i}\wedge \delta
\Sigma^{\va}_{i} ) + \frac{1}{2  } \int_{S} (\delta \e_i\wedge\delta \e^i)\,.
\ee
Within this extended phase space, the translation generator \eqref{Pb} becomes differentiable. However, this spoils the Hamiltonian nature of the electric generator \eqref{Gb}. In order to remedy this, it is necessary to add a boundary term to \eqref{Gb} in terms of the boundary co-frame field. The requirement that the Hamiltonian generator vanishes on-shell dictates the form of such boundary term. More precisely, if we demand the boundary simplicity constraint 
\be\la{simp}
S^i:= \Sigma_i-\frac12 (\e\times \e)_i\myeq{S}0
\ee
to hold\footnote{This can be equivalently written as the constraint that the boundary co-frame equals the pull-back of  the bulk frame field $e^i \myeq{S}\e^i$.},
then the canonical generators for the  gauge transformations associated with the electric and translational constraints \eqref{G-cons}, \eqref{P-cons} are given by:
\begin{align}
&G_\alpha=-\int_{\cB} \rd_A\alpha^i\wedge \Sigma_i
+ \tfrac12 \int_S \alpha^i (\e\times \e)_i
\,, &\text{Electric gauge}\la{G}\\
&P_\varphi=-\int_{\cB} \rd_A \varphi^i\wedge \rd_A e_i { + { \int_S \varphi^i \rd_A(e_i-\e_i)}
}\,,&\text{Spatial translation}\la{P}\,.
\end{align}
We see that with the introduction of the boundary electric edge modes $\e^i$, 
$G_\alpha$ vanishes on-shell even for parameters $\alpha_i$ that are non-vanishing on $S$; explicitly,
\be
\label{eqn:defGint}
G_\alpha=\int_{\cB} \alpha^i \rd_A \Sigma_i - \int_S \alpha^i S_i\,\hat{=}\, 0\,, 
\ee
where $\hat =$ denotes on-shell of $\eqref{simp}$, or equivalently $s^i:=e^i-\e^i\myeq{S}0$.
This is in agreement with the general philosophy of \cite{Donnelly:2016auv} where it is shown that the restoration of boundary gauge symmetry goes hand-in-hand with the introduction of edge modes.

On the other hand, the translation Hamiltonian generator \eqref{P} does not vanish on-shell in general for translations that do not vanish on the boundary; explicitly, 
\be
P_\varphi=\int_{\cB} \varphi^i \rd_A P_i{ -{ \int_S \varphi^i \rd_A\e_i}}\,\hat{=}\, - \int_S \varphi^i P_i\,,
\ee 
where we have used the translation constraint $\rd_A P_i\simeq0$.
In \cite{Freidel:2019ees} the boundary condition $P^i\myeq{S_P}0$ was assumed, where $S_P$ is a punctured boundary sphere. This means that non-zero symmetry charges  were associated only to the punctures, where
the source of momenta were located. While we are not going to introduce distributional curvature and momentum at the punctures here, as we are simply interested in the classical algebra of the gravitational kinematical charges, one can think of \eqref{P} as the generator of gauge translation in the bulk and on the boundary at the locus  of vanishing $P^i$, and as a generator of  boundary symmetry otherwise.

Before looking in more detail at the symmetry algebra, let us conclude this part with an interesting observation concerning the momentum $P_i$ \eqref{P} which plays a key role in the symplectic structure. $P_i$ is simply the canonical momentum conjugate to the co-frame $e^i$. In fact, assuming the simplicity constraint in the form $e=\e$ on the boundary, and denoting $\omega$ the integrand of the symplectic structure,  we have:
\ba
\omega
&=& \delta A^i \wedge \delta \Sigma_i + \tfrac12\rd (\delta e_i\wedge \delta e^i)\cr
&=&\delta A^i \wedge (e\times \delta e)_i + \tfrac12 \rd_A \delta e_i \wedge \delta e^i 
- \tfrac12 \delta e^i \wedge \rd_A \delta e_i  \cr
&=& (\delta A \times e)_i \wedge \delta e^i + \rd_A \delta e_i \wedge \delta e^i\cr
&=& \delta P_i\wedge \delta e^i.\la{symp2}
\ea
This shows that the bulk plus boundary phase space $(A^i,\Sigma_i, \e^i)$, supplemented by the matching  condition $e=\e$ on the boundary, can be   expressed equivalently in terms of the canonical pair $(P_i,e^i)$.

 In section \ref{sec:gaugetransf}, we   establish explicitly that  the generators \eqref{G}, \eqref{P} give respectively the electric and the translational gauge transformations. As a first step, we introduce a useful duality map between $\su(2)$-valued 1-forms and $\su(2)$-valued 2-forms below in section \ref{sec:dualitymap}.

\subsection{A duality map}
\label{sec:dualitymap}

In order to write the phase space field transformations in a symplectic manner, we need to introduce the following map
from Lie algebra valued 2-forms to Lie algebra valued 1-forms:
\be
\begin{array}{lrcl}
\rho\,: &\Omega_2(\su(2)) &\longrightarrow& \Omega_1(\su(2))
\\
&  B^i &\longmapsto& \rho(B)_{i}:=\tilde{B}_i
\end{array}
\qquad\textrm{such that}\quad
(\tilde{B}\times e)_i= B_i \,.
\ee
This ``tilde'' map is the inverse of $ A^i\to (A\times e)^i$ from $\Omega_1(\su(2)) \to \Omega_2(\su(2))$.
It exists as long as $e$ is invertible.
An explicit formula can be given. In order to do so,  we expand the forms in the corresponding basis
\be
B_i= B_i{}^j \Sigma_j,\qquad 
 \tilde{B}^i= \tilde{B}^i{}_j e^j. 
\ee
The components of the forms are related using shifts by their traces $B = B_i{}^i$ and $\tilde{B} = \tilde{B}{}^i{}_{i}$:
\be
B_i{}^j = \tilde{B}\delta_i^j- \tilde{B}^j{}_i,\qquad 
{\tilde{B}}^j{}_i = \frac{{B}}{2}\delta^j_i -B_i{}^j
\,.
\ee
The trace $B = B_i{}^i$ appears naturally when evaluating the 3-form
\be
B_i \wedge e^i = {2} \tilde{B}^i \wedge \Sigma_i =  B\,  \mathrm{det}(e),
\ee
where $ \mathrm{det}(e) = e^1\wedge e^2 \wedge e^3$ is the volume form\footnote{One uses that  
\be
e^i\wedge \Sigma_j=\delta^i_j \mathrm{det}(e).
\ee.}.
This duality map $B^i\to\tilde{B}_i$ defines a symmetric product on $\Omega_2(\su(2))$
since 
\be
\tilde{A}^i\wedge B_i = A_i \wedge \tilde{B}^i =(\tilde{A} \tilde{B} - \tilde{A}^i{}_j  \tilde{B}^j{}_i ) \,  \mathrm{det}(e)
\,.
\ee

Examples of the value of the map on some relevant $\su(2)$-valued 2-forms include
\be
\widetilde{\Sigma}^i = \frac1{2} e^i\,,\qquad
\widetilde{(\alpha\times \Sigma)}{}^i =(\alpha \times e)^i\,,\qquad
\widetilde{P}{}^i=\widetilde{\rd_A e}{}^i= \gamma {K^i}
\,. 
\ee
Another useful identity that follows  from $(\varphi \times B)^i= (\iota_{\hat\varphi}B\times e)^i +  \iota_{\hat\varphi} (e\times B)^i$, where $\hat\varphi^i =\iota_{\hat \varphi }e^i$, is 
\be \la{id}
\widetilde{(\varphi \times B)}^i= \iota_{\hat\varphi} B^i
+\stackon[-8pt]{$\iota_{\hat \varphi }  (e\times B)^i$.}{\vstretch{1.5}{\hstretch{2.4}{\widetilde{\phantom{\;\;\;\;\;\;\;}}}}}
\ee

\subsection{Gauge transformations}
\label{sec:gaugetransf}

\paragraph{Electric gauge transformations:}

The electric gauge transformations  generated by the constraint $G_{\alpha}$ given in \eqref{G} act on the  bulk and boundary fields in the following way:
\begin{align}
& \delta^{\e}_{\alpha}\Sigma_i=(\alpha\times \Sigma)_i\,,
\qquad \delta^{\e}_{\alpha}A^i=-\rd_A \alpha^i\,,\qquad
\delta^{\e}_{\alpha} \e^i=(\alpha\times \e)^i\,.
\la{delta-G}
\end{align}
This in turn implies that
\be
\delta^{\e}_{\alpha} e^i=(\alpha\times e)^i
\,\qquad
\delta^{\e}_{\alpha}  P_i= (\alpha\times P)_i,\qquad
\delta^{\e}_{\alpha}  F_i= (\alpha\times F)_i
.
\ee
In particular, $e$ and $\e$ transform in the same way, so that the simplicity constraint is preserved under electric gauge transformations.
We also see that all the charge aspects $(\Sigma, P, F)$ transform in the same manner.

In order to derive these gauge transformations, one  evaluates the extended pre-symplectic form $\Omega$ given in \eqref{Oext} on the Hamiltonian vector field generated by the constraint $G_{\alpha}$:
\ba
\la{deltaG}
I_{\delta^{\e}_\alpha}\Omega
 &=& 
 - \int_\cB \rd_A\alpha^i\wedge \delta \Sigma_i  
-\int_{\cB} \delta A^i\wedge (\alpha \times \Sigma)_i 
+ \int_S (\alpha\times \e)_i\wedge \delta \e^i\cr
&=&- \int_{\cB} (\delta A\times \alpha)^i \wedge  \Sigma_i - \int_\cB \rd_A\alpha^i\wedge \delta \Sigma_i + \int_S \alpha^i  (\e \times \delta \e)_i
=
\delta G_{\alpha}
\,.
\ea
This establishes (\ref{delta-G}).

\bigskip

\paragraph{Translational gauge transformations:}

The translational  transformations  generated by the constraint $P_{\vphi}$ given in \eqref{P}  act on the  bulk and boundary fields in the following way:
\ba
\la{delta-P}
& \delta^{\rt}_{\varphi}\Sigma_i
=\rd_A(\varphi\times e)_i
\,,\qquad \delta^{\rt}_{\varphi}A^i=\widetilde{( \varphi\times F)}^i\,,\qquad
&\delta^{\rt}_{\varphi}\e^i\myeq{S} \rd_A \varphi^i
\,.
\ea
These relations imply that the action of the translations takes the same form on all the electric, momentum and curvature aspects :
\be\label{delta-P1}
\delta^{\rt}_{\varphi}\Sigma_i=\rd_A \widetilde{(\varphi\times \Sigma)}_i,\qquad
\delta^{\rt}_{\varphi} P_i= \rd_A \widetilde{(\varphi\times P)}_i,\qquad
\delta^{\rt}_{\varphi} F_i=\rd_A \widetilde{(\varphi\times F)}_i.
\ee

In order to show that these expression are gauge transformations 
canonically generated by  $ P_\varphi $, we evaluate the variation of the momentum constraint:
\ba
\delta P_\varphi&=&-\int_{\cB} (\delta A\times  \varphi)^i\wedge \rd_A e_i
-\int_{\cB} \rd_A \varphi^i\wedge (\delta A\times e)_i
-\int_{\cB} \rd_A \varphi^i\wedge \rd_A \delta e_i + { \int_{S} \rd_A \varphi^i\wedge  \delta (\e_i-e_i)} 
- \int_{S} \delta A^i  \wedge (\varphi \times s)_i   \cr
&=&-\int_{\cB} \delta A_i \wedge ( \varphi \times  \rd_A e)_i
- \int_{\cB}  \delta A_i \wedge (\rd_A \varphi \times e)^i
-\int_{\cB} (F\times \varphi)_i\wedge  \delta e^i 
+ \int_{S} \rd_A \varphi_i\wedge  \delta \e^i- \int_{S} \delta A^i  \wedge (\varphi \times s)_i   \,,
\ea
where we have denoted the boundary simplicity constraint by $s^i:=e^i-\e^i$.
Using the duality map, we can now write the  variation of $P_\varphi$ as 
\ba
\delta P_\varphi 
 &=& 
- \int_{\cB}  \delta A_i \wedge \rd_A ( \varphi \times e)^i
-\int_{\cB} \widetilde{(F\times \varphi)}_i\wedge (e\times \delta e)^i  
+ \int_{S} \rd_A \varphi^i\wedge  \delta \e_i  - \int_{S} \delta A^i  \wedge (\varphi \times s)_i    \n\\
&=& 
- \int_{\cB}  \delta A_i \wedge \rd_A ( \varphi \times e)^i
+\int_{\cB} \widetilde{(\varphi \times F )}_i\wedge \delta\Sigma^i  
+ \int_{S} \rd_A \varphi^i\wedge  \delta \e_i - \int_{S} \delta A^i  \wedge (\varphi \times s)_i    \cr
&=& I_{\delta^{\rt}_\varphi} \Omega - \int_{S} \delta A^i  \wedge (\varphi \times s)_i   
\,,
\la{deltaP}
\ea
where in the last equality we have used  the action of the translations given  in \eqref{delta-P}.
This shows that $P_\varphi $  is the generator of translation \eqref{delta-P} if the boundary Gauss law $s_i=0$ is satisfied.
From the transformation of the flux 2-form $\Sigma$ we deduce the transformation of the frame field $e$:
\be
{ \delta^\rt_\varphi e^i = \rd_A \varphi^i +\widetilde{(\varphi\times P)}^i.}
\la{deltae}
\ee
This shows that the translation acts on the momenta as follows:
\ba
\delta^\rt_\varphi P_i=  \delta^\rt_\varphi (\rd_A e)_i 
 &=& (\delta^\rt_\varphi A \times e)_i
 + \rd_A \delta^\rt_\varphi e_i = (\varphi\times F) + \rd_A^2 \varphi^i +\rd_A \widetilde{(\varphi\times P)}_i=\rd_A \widetilde{(\varphi\times P)}_i\,,
\ea
which allows us to recover the transformation law for the momentum $P$ anticipated in \eqref{delta-P1}.

\medskip

As one can see from above by comparing the action of the translation given above in \eqref{delta-P} and \eqref{deltae}, the bulk frame field $e$ does not transform in the same way as the edge mode $\e$ on the boundary. This means that boundary translations do not generally preserve the boundary simplicity constraint.
More precisely, we find  that 
\be
{ \delta^\rt_\varphi S_i = (\rd_A\varphi\times s)_i + (\varphi\times P)_i, \qquad \delta^\rt_\varphi s^i = \widetilde{(\varphi\times P)}^i\,.}
\ee
 As we will see in the next section, this puzzling feature does not  affect the computation of the charge algebra, but it simply means that the charge algebra only closes  on the support of the boundary constraint $s^i=0$.

\subsection{Translation versus Diffeomorphism}

Now that we have derived the expressions of the translational gauge transformations, we have to compare them to the diffeomorphisms.
First, using the expressions \eqref{delta-P}, and the identity (\ref{id}), we can relate the action of translations to the action of the covariant diffeomorphisms given by the covariant Lie derivative $L_{\hat{\varphi}}:=\iota_{\hat{\varphi}}\rd_A + \rd_A\iota_{\hat{\varphi}} $~:
\ba
&\delta^{\rt}_{\varphi}\e^i= L_{\hat{\varphi}} \e^i - \iota_{\hat{\varphi}}P^i
\,,\quad \delta^{\rt}_{\varphi}\Sigma_i
=L_{\hat{\varphi}} \Sigma_i - \iota_{\hat{\varphi}}(\rd_A \Sigma)_i
\,,\quad \delta^{\rt}_{\varphi}A^i=\iota_{\hat{\varphi}} F^i
-\stackon[-8pt]{$\iota_{\hat \varphi }  (\rd_A P^i)$}{\vstretch{1.5}{\hstretch{2.4}{\widetilde{\phantom{\;\;\;\;\;\;\;}}}}}
\,.
\la{delta-T}
\ea
We see that translations induce an extra-term besides the Lie derivative, which vanishes on-shell for both the flux 2-form $\Sigma$ and the Ashtekar-Barbero connection $A$.

Diffeomorphism can actually be understood as field dependent translations. In order to show this, let us start with the covariant diffeomorphism generator defined by
\be\la{D}
D_{\hat{\varphi}} =-\int_{\cB}\iota_{\hat \varphi} \Sigma^i\wedge F_i -\frac12 \int_S  \e^i \wedge L_{\hat \varphi}\e_i\,.
\ee
This operator generates the covariant diffeomorphism symmetry $\delta^{\rd}_{\hat\varphi}$ \footnote{$\delta^{\rd}_{\hat\varphi}$ is equivalent on-shell to the action of the covariant Lie derivative. More precisely we have that 
\be
(L_{\hat{\varphi}}-\delta_{\hat\varphi}^{\rd})\Sigma_i = \imath_{\hat\varphi}\rd_A\Sigma_i,\qquad
(L_{\hat{\varphi}}-\delta_{\hat\varphi}^{\rd})A^i = 0,\qquad
(L_{\hat{\varphi}}-\delta_{\hat\varphi}^{\rd})\e^i = 0.
\ee }
\be
 \delta^{\rd}_{\hat\varphi}\Sigma_i
=\rd_A\imath_{\hat\varphi} \Sigma_i
\,,\qquad \delta^{\rd}_{\hat\varphi}A^i=\iota_{\hat\varphi} F^i(A)\,,\qquad
\delta^{\rd}_{\hat\varphi}\e^i= L_{\hat\varphi} \e^i
\,.\la{delta-D} 
\ee
We derive these diffeomorphism transformation properties \eqref{delta-D} by evaluating the extended pre-symplectic form $\Omega$ on the constraint $D_{\hat\varphi}$ and using that the boundary value of   $\hat\varphi $ is tangent to $S$ :
\ba
\delta D_{\hat{\varphi}} &=&{
-\int_{\cB}\iota_{\hat \varphi} \delta \Sigma_i \wedge F_i
-\int_{\cB}\iota_{\hat \varphi} \Sigma^i\wedge \rd_A \delta A_i 
{ -\int_S (\delta A^i\times   \iota_{\hat{\varphi}}\e_i )\wedge \e^i } - \int_S  \delta \e^i \wedge L_{\hat \varphi}\e_i,}\n \\
&=&
\int_\cB  \iota_{\hat{\varphi}}F_i\wedge  \delta \Sigma_i
-\int_\cB \delta A^i\wedge \rd_A \iota_{\hat{\varphi}}\Sigma_i 
{ -\int_S \delta A^i\wedge  \iota_{\hat{\varphi}}S_i } + \int_S   L_{\hat \varphi}\e_i\wedge \delta \e^i\cr
&=& I_{\delta^{\rt}_\varphi} \Omega { -\int_S \delta A^i\wedge  \iota_{\hat{\varphi}}S_i },\la{deltaD}
\ea
from which we can read the  transformations (\ref{delta-D}).
We have used that $S_i =\Sigma_i-\tfrac12(e\times e)_i$.

We can then compute  the difference between translations and diffeomorphisms:
 \be
 \la{delta-DP} 
(\delta^{\rd}_{\hat\varphi}-\delta^{\rt}_\varphi) \Sigma_i
= 0
\,,\quad (\delta^{\rd}_{\hat\varphi}-\delta^{\rt}_\varphi)A^i=
\stackon[-8pt]{$\iota_{\hat \varphi }  (\rd_A P^i)$}{\vstretch{1.5}{\hstretch{2.4}{\widetilde{\phantom{\;\;\;\;\;\;\;}}}}}
\,,\quad 
(\delta^{\rd}_{\hat\varphi}-\delta^{\rt}_\varphi)\e^i= \iota_{\hat\varphi} P_i
\,.
\ee
As expected, the difference between transformations vanishes on-shell (when $\rd_{A}\Sigma=\rd_{A}P=0$) for the bulk variables.
The key difference between the translation and diffeomorphism shows up in their action on the boundary variable $\e$. 
This  in turns implies  that the diffeomorphism generator always  preserves the boundary simplicity constraints, namely  $\delta^{\rd}_{\hat\varphi} s^i=0$, even if the translations do not.

Finally,  it is straightforward to show that the difference between a field dependent  translation and the diffeomorphism generator vanishes on-shell of the boundary simplicity constraints.
\be
D_{\hat{\varphi}}\hat{=} P_{\iota_{\hat{\varphi}} e}.
\ee

\section{Charge algebra}
\label{sec:algebra}

Now that we have established the Hamiltonians \eqref{G}, \eqref{P} as the canonical generators of electric and translational gauge symmetry, we can study the algebra of the Hamiltonian charges associated to these gauge symmetries. We rely on the expression of the Poisson bracket between two Hamiltonian generators in terms of the generators variations, namely
\be
\{H_\alpha,H_{\beta} \} = \delta_{\beta}H_{\alpha}  \,.
\ee
Moreover, since the generators of gauge transformations $\delta_\alpha=\{H_\alpha,\cdot\}$ form a closed algebra
$[\delta_\alpha,\delta_\beta]=-\delta_{[\alpha,\beta]}$, the Jacobi identity implies  the consistency condition $\{H_\alpha,H_\beta\}= H_{[\alpha,\beta]}+ c(\alpha,\beta)$ where $c(\alpha,\beta)$ is central. 

We can thus analyze  the various sectors of the algebra formed by the generators $G_{\alpha}$ and $P_{\vphi}$ by means of the transformation properties under electric gauge transformations \eqref{delta-G} and under translations \eqref{delta-P} derived in the previous section.
This will lead us to show that the constraints $G_{\alpha}$ and $P_{\vphi}$ form an $\mathfrak{isu}(2)$ Poincar\'e Lie algebra.

\subsection{Electric-Electric sector}

The electric-electric sector consists of the Poisson brackets of the electric gauge  generators $G$ with each other.
This is the simplest case, as it is immediate to check that using the electric gauge transformations \eqref{delta-G}, we obtain the following brackets:
\ba
\{G_\alpha,G_\beta\}&=&\delta^\e_\beta G_\alpha=
- \int_{\cB} (\delta_\beta A\times \alpha)^i \wedge  \Sigma_i - \int_\cB \rd_A\alpha^i\wedge \delta_\beta \Sigma_i + \int_S \alpha^i  (\e \times \delta_\beta \e)_i\cr
&=& \int_{\cB} (\rd_A\beta \times \alpha)^i \wedge  \Sigma_i - 
\int_\cB \rd_A\alpha^i\wedge (\beta\times \Sigma)_i + \int_S \alpha^i  (\e \times (\beta\times \e))_i\cr
&=&
-\int_{\cB} \rd_A (\alpha\times\beta)^i\wedge \Sigma_i
+\frac{1}{2}\int_S(\alpha\times\beta)^i (\e\times \e)_i \n\\
&=&
G_{(\alpha\times \beta)}\,.
\ea
The electric-electric sector thus closes and we recover the expected $\su(2)$  Lie algebra, as for the flux observables in loop quantum gravity.

\subsection{Electric-Translation sector}

The electric-translation sector consists of the Poisson brackets of the electric gauge  generators $G_{\alpha}$ with the translation generators $P_{\vphi}$.
For this mixed sector we start with the definition of the electric constraint \eqref{G} and compute its variation under translations using \eqref{delta-P}. After a sequence of elementary operations involving the duality map, the cross product and integrations by part, we obtain:
\ba
\{   G_{\alpha},P_\varphi\}&=&\delta^\rt_\varphi G_{\alpha}=
-\int_{\cB}  (\delta^\rt_\varphi A\times   \alpha)_i\wedge \Sigma^i
-\int_{\cB}  \rd_A \alpha^i\wedge \delta^\rt_\varphi  \Sigma_i
+\int_S \alpha^i (\delta^\rt_\varphi \e \times \e)_i
\n\\
&=&-\int_{\cB}  \widetilde{( \varphi\times F)}^i\wedge ( \alpha\times \Sigma)_i
-\int_{\cB}  \rd_A \alpha^i\wedge \rd_A ( \varphi \times e)^i 
+\int_S \alpha^i (\rd_A \varphi  \times \e)_i
\n\\
&=&-\int_{\cB}  ( \varphi \times  F )_i\wedge  \widetilde{( \alpha\times \Sigma)}^i
-\int_{\cB}  \rd^2_A \alpha_i\wedge  ( \varphi \times e)^i
+\int_S \rd_A \alpha_i\wedge  ( \varphi \times e)^i 
+\int_S (\alpha \times \rd_A \varphi)_i \wedge   \e^i
\n\\
&=&\int_{\cB}  (F\times \varphi) \wedge  ( \alpha\times e)_i
-\int_{\cB} (( F\times \alpha)\times \varphi )_i\wedge e^i
+\int_S \rd_A (\alpha \times \varphi)_i \wedge e ^i 
+ {\int_S (\alpha \times \rd_A \varphi)_i \wedge   (\e^i-e^i)}
\n\\
&=&\int_{\cB} (( F\times (\varphi\times \alpha ))_i\wedge e^i
+\int_S \rd_A (\alpha \times \varphi)_i \wedge e ^i + {\int_S (\alpha \times \rd_A \varphi)_i \wedge   (\e^i-e^i)}\cr
&=&{ -\int_{\cB} \rd_A(\alpha \times \varphi)^i\wedge \rd_A e_i - {\int_S \rd_A (\alpha \times  \varphi)_i \wedge   (e^i-\e^i)} }+ {\int_S (\rd_A\alpha \times  \varphi)_i \wedge  (e^i-\e^i)}\,, \cr
&=& { P_{(\alpha\times \varphi)}} + {\int_S (\rd_A\alpha \times  \varphi)_i \wedge   s^i}
\,.
\ea
The extra-term vanishes on-shell of the boundary Gauss Law $s^i=e^i-\e^i\overset{S}=0$.
Thus, assuming the boundary simplicity $e^i\overset{S}=\e^i$, we recover the expected Poisson bracket, namely
\be\la{GP}
\{  G_{\alpha},P_\varphi \}\hat ={ P_{(\alpha\times \varphi)}}\,.
\ee

Instead of starting with the generator $G_{\alpha}$ and computing its variation under translations, we can do the reverse and start with the generator $P_{\vphi}$ as defined in \eqref{P} and compute its variation under electric gauge transformations using \eqref{delta-G}. This allows to check that the gauge transformations  \eqref{delta-G}, \eqref{delta-P} are indeed consistent with the antisymmetry of the Poisson bracket, namely that 
\be
\delta^\rt_\varphi G_{\alpha}=-\delta^\e_{\alpha} P_\varphi\,.
\ee
So, in order to check this,  we plug in the electric variations  \eqref{delta-G} in the general expression  \eqref{deltaP} for the variation of the translation generator $\delta P_{\vphi}$ and compute:
\ba
\{   P_\varphi, G_{\alpha} \}&=&\delta^\e_{\alpha} P_\varphi =
- \int_{\cB}  \delta^\e_{\alpha} A_i \wedge \rd_A ( \varphi \times e)^i
+\int_{\cB} \widetilde{(\varphi \times F )}_i\wedge \delta^\e_{\alpha} \Sigma^i  
+ \int_{S} \rd_A \varphi^i\wedge \delta^\e_{\alpha} \e_i - \int_{S} \delta^{\e}_\alpha A^i  \wedge (\varphi \times s)_i  \n\\
&=& \int_{\cB}  \rd_A\alpha^i \wedge \rd_A( \varphi  \times e)_i
+\int_{\cB} (\varphi \times F )_i \wedge (\alpha \times  e)^i + { \int_{S} \rd_A \varphi^i\wedge  (\alpha \times \e)_i} + \int_{S} \rd_A\alpha^i  \wedge (\varphi \times s)_i  \n\\
&=& \int_{\cB}  (F\times \alpha)_i \wedge ( \varphi  \times e)^i
-\int_{\cB} (F\times \varphi  )_i \wedge (\alpha \times  e)^i- \int_{S}  \rd_A\alpha^i \wedge ( \varphi  \times \e)_i + { \int_{S} (\rd_A \varphi\times\alpha)_i \wedge  \e^i}\n\\
&=& \int_{\cB}  (\varphi \times \alpha)_i \wedge ( F \times e)^i
+ { \int_{S} \rd_A (\varphi\times \alpha)_i \wedge \e^i}\n\\
&=& -\int_{\cB}  \rd_A(\varphi\times \alpha)_i \wedge \rd_A e^i + \int_{S}  (\varphi \times \alpha)_i \wedge  \rd_A (e-\e)^i \n\\
&=&{ P_{(\varphi\times \alpha)} } \,.
\ea
%
This shows that the electric and translation transformations are indeed consistent when $s^i=0$. And we have established that the electric-translation sector closes when the boundary simplicity constraint is satisfied, in which case the bracket  \eqref{GP} holds.

\subsection{Translation-Translation sector}

Finally, we turn to the translation-translation sector. We now want to prove that the translation generators commute on-shell, that is we want to establish that
\be 
\{P_{\xi}, P_{\varphi}\} \simeq0
\,,
\ee
where the equality $ \simeq$ means we are on-shell of the three   conservation laws  \eqref{G-cons},  \eqref{P-cons}, \eqref{F-cons}.
Although this is the simplest commutation relation in the end, it turns out that this is the most involved evaluation.
We start by plugging the transformations \eqref{delta-P} into \eqref{deltaP} which, together with the identity \eqref{id} for the duality map,
yields when $s^i=e^i-\e^i\overset{S}=0$:
\ba
\{P_{\xi}, P_{\varphi}\}&=& \delta^\rt_{\varphi}P_{\xi} \hat =
- \int_{\cB}  \widetilde{( \varphi\times F)}_i \wedge \rd_A ( \xi \times e)^i
+\int_{\cB} \widetilde{(\xi \times F )}_i\wedge \rd_A ( \varphi \times e)^i 
+  \int_{S} \rd_A \xi^i\wedge   \rd_A \varphi_i  
\n\\
&{=}&- \int_{\cB}  \iota_{\hat\varphi} F_i \wedge \rd_A ( \xi \times e)^i
+\int_{\cB} \iota_{\hat \xi} F_i\wedge \rd_A ( \varphi \times e)^i 
-  \int_{S}  ( \varphi\times \xi)^i F_i \n\\
&-& \int_{\cB}\stackon[-8pt]{$\iota_{\hat \varphi }  (e\times F)^i$}{\vstretch{1.5}{\hstretch{2.4}{\widetilde{\phantom{\;\;\;\;\;\;\;}}}}} \wedge \rd_A ( \xi \times e)^i
+\int_{\cB}\stackon[-8pt]{$\iota_{\hat \xi  }  (e\times F)^i$}{\vstretch{1.5}{\hstretch{2.4}{\widetilde{\phantom{\;\;\;\;\;\;\;}}}}} \wedge \rd_A ( \varphi \times e)^i\,,
\la{PP}
\ea
where we have  integrated by parts on the sphere and introduced vectors $(\hat{\varphi}, \hat{\xi})$ related to the translation gauge parameters $(\varphi^i,\xi^i)$  via $\imath_{\hat\varphi} e^i = \varphi^i$.

Let us focus on the term
\ba
  \iota_{\hat \varphi} F_i \wedge \rd_A ( \xi \times e)^i&=& -F_i \wedge \iota_{\hat \varphi} \rd_A \iota_{\hat \xi} \Sigma^i \,.
\ea
The key commutation of Cartan calculus, $[L_{\hat\xi} ,\iota_{\hat\varphi} ]=\iota_{[\hat\xi, \hat\varphi]}$
between Lie derivative and interior product,  implies the identity
\be\label{Cartan}
\iota_{\hat \xi} \rd_A \iota_{\hat \varphi}-\iota_{\hat \varphi} \rd_A \iota_{\hat \xi}=\iota_{[\hat\xi, \hat\varphi]} -\rd_A(\iota_{\hat \xi}\iota_{\hat \varphi})-\iota_{\hat \xi}\iota_{\hat \varphi}\rd_A.
\ee
This allows to compute:
\ba
\iota_{\hat\varphi} F_i \wedge \rd_A ( \xi \times e)^i
- \iota_{\hat\xi} F_i\wedge \rd_A ( \varphi \times e)^i &=& F_i \wedge (\iota_{\hat \xi} \rd_A \iota_{\hat \varphi}-\iota_{\hat \varphi} \rd_A \iota_{\hat \xi} )\Sigma^i,\cr
&=& F_i \wedge (\iota_{[\hat \xi,\hat \varphi]}  +\iota_{\hat \varphi}  \iota_{\hat \xi}\rd_A -\rd_A \iota_{\hat{\xi}} \iota_{\hat \varphi})\Sigma^i.
\ea
We thus obtain
\ba
\{P_{\xi}, P_{\varphi}\}&\hat =&
 \int_{\cB}  \iota_{[\hat\xi, \hat\varphi]}F_i \wedge \Sigma^i
  { -} \int_{\cB}F^i\wedge  \rd_A(\iota_{\hat \varphi}\iota_{\hat \xi}\Sigma _i) 
  { -} \int_{\cB}F^i\wedge  \iota_{\hat \varphi}\iota_{\hat \xi} \rd_A\Sigma _i
  { -}\int_{S}  ( \varphi\times \xi)^i F_i \n\\
 &-& \int_{\cB}\stackon[-8pt]{$\iota_{\hat \varphi }  (e\times F)^i$}{\vstretch{1.5}{\hstretch{2.4}{\widetilde{\phantom{\;\;\;\;\;\;\;}}}}} \wedge \rd_A ( \xi \times e)^i
+\int_{\cB}\stackon[-8pt]{$\iota_{\hat \xi  }  (e\times F)^i$}{\vstretch{1.5}{\hstretch{2.4}{\widetilde{\phantom{\;\;\;\;\;\;\;}}}}} \wedge \rd_A ( \varphi \times e)^i\n\\
 &=& \int_{\cB} F_i \wedge  \iota_{[ \hat\varphi, \hat\xi]}\Sigma^i
 { +}\int_{\cB}\rd_A F^i  (\iota_{\hat \varphi}\iota_{\hat \xi}\Sigma _i) 
   { -} \int_{\cB} \iota_{\hat \varphi}\iota_{\hat \xi} F^i   \rd_A\Sigma _i\n\\
   &-& \int_{\cB}\stackon[-8pt]{$\iota_{\hat \varphi }  (e\times F)^i$}{\vstretch{1.5}{\hstretch{2.4}{\widetilde{\phantom{\;\;\;\;\;\;\;}}}}} \wedge \rd_A ( \xi \times e)^i
+\int_{\cB}\stackon[-8pt]{$\iota_{\hat \xi  }  (e\times F)^i$}{\vstretch{1.5}{\hstretch{2.4}{\widetilde{\phantom{\;\;\;\;\;\;\;}}}}} \wedge \rd_A ( \varphi \times e)^i\n\\
 &=&- \int_{\cB} (\iota_{[ \hat\varphi, \hat\xi]}e^i)  \rd_A P_i
  { +}\int_{\cB} (\iota_{\hat \varphi}\iota_{\hat \xi}\Sigma _i) \rd_A F^i 
  { -} \int_{\cB} (\iota_{\hat \varphi}\iota_{\hat \xi} F^i)   \rd_A\Sigma _i\n\\
   &+& \int_{\cB}\stackon[-8pt]{$\iota_{\hat \varphi } \rd_A P^i$}{\vstretch{1.5}{\hstretch{2.4}{\widetilde{\phantom{\;\;\;\;\;\;\;}}}}} \wedge \rd_A ( \xi \times e)^i
-\int_{\cB}\stackon[-8pt]{$\iota_{\hat \xi  }  \rd_A P^i$}{\vstretch{1.5}{\hstretch{2.4}{\widetilde{\phantom{\;\;\;\;\;\;\;}}}}} \wedge \rd_A ( \varphi \times e)^i
\n\\
&\simeq&0\,.
\ea
This expression vanishes on-shell, i.e. when assuming the constraints and Bianchi identity, $\rd_{A}\Sigma=\rd_{A}P=\rd_{A}F=0$. This concludes our proof that the electric and translational charges form indeed a closed Poincar\'e algebra.

\subsection{Diffeomorphisms algebra}

As we have checked that the electric and momentum constraints, with the appropriate boundary terms, form a Poincar\'e Lie algebra on-shell, it is interesting to turn back to the diffeomorphisms. We come back to the expression \eqref{D} of the diffeomorphism constraints as field dependent translations, and we would like to verify that we recover the correct diffeomorphisms algebra.

Plugging the diffeomorphism transformations \eqref{delta-D} in the general variation of the diffeomorphism generators \eqref{deltaD}, we can compute the bracket
\ba
\{D_{\hat{\xi}},D_{\hat{\varphi}} \}=\delta^\rd_{\hat \varphi } D_{\hat{\xi}} &=& 
\int_\cB\rd_A \iota_{\hat\varphi}\Sigma_i\wedge \iota_{\hat{\xi}}F_i
-\int_\cB\iota_{\hat\varphi} F^i \wedge \rd_A \iota_{\hat{\xi}}\Sigma_i
{ -\int_S \iota_{\hat{\varphi}}F^i \wedge   (\iota_{\hat{\xi}}\e \times  \e)_i }- {\int_S L_{\hat \varphi} \e^i \wedge L_{\hat \xi} \e_i}\n\\
&=&
 \int_\cB (  \iota_{\hat\varphi}   \rd_A \iota_{\hat{\xi}}\Sigma_i  -\iota_{\hat\xi}   \rd_A \iota_{\hat{\varphi}}\Sigma_i)\wedge F^i- \int_S(  \rd_A  \iota_{\hat{\varphi}} \e_i \wedge \rd_A  \iota_{\hat{\xi}} \e_i 
+ \iota_{\hat{\xi}} \iota_{\hat{\varphi}}  \rd_A \e_i\wedge  \rd_A \e_i)\n\\
&-&\int_S(\iota_{\hat\xi}   \rd_A \iota_{\hat{\varphi}} \e^i-\iota_{\hat\varphi}   \rd_A \iota_{\hat{\xi}} \e^i)\wedge \rd_A \e_i { -\tfrac12 \int_S \iota_{\hat{\xi}}\iota_{\hat{\varphi}} F^i \wedge   (\e \times  \e)_i }\,.
\ea
We now use again the Cartan calculus identity (\ref{Cartan}) to obtain
\ba
\{D_{\hat{\xi}},D_{\hat{\varphi}}  \}&=&
-\int_\cB \iota_{[\hat\xi,\hat\varphi]}\Sigma_i  \wedge F^i
+\int_\cB\iota_{\hat \xi}\iota_{\hat \varphi}\rd_A\Sigma_i  \wedge F^i
 { +}\int_\cB \rd_A(\iota_{\hat \xi}\iota_{\hat \varphi}\Sigma_i ) \wedge F^i \n\\
&-& \int_S \iota_{[\hat\xi,\hat\varphi]} e_i   \rd_A e_i 
-\int_S \rd_A \varphi^i\wedge \rd_A \xi_i { -\tfrac12 \int_S \iota_{\hat{\xi}}\iota_{\hat{\varphi}} F^i \wedge   (\e \times  \e)_i }\n\\
& =& D_{[\hat\xi,\hat\varphi]}
+\int_\cB\rd_A\Sigma_i   \iota_{\hat \xi}\iota_{\hat \varphi}F^i
-\int_\cB\iota_{\hat \xi}\iota_{\hat \varphi}\Sigma_i   \rd_A F^i 
+\int_S \iota_{\hat \xi}\iota_{\hat \varphi}\Sigma_i   F^i
+\int_S(\xi\times \varphi)^i F_i { -\tfrac12 \int_S \iota_{\hat{\xi}}\iota_{\hat{\varphi}} F^i \wedge   (\e \times  \e)_i }\n\\
& =& D_{[\hat\xi,\hat\varphi]}
+\int_\cB\iota_{\hat \xi}\iota_{\hat \varphi}  F^i \rd_A\Sigma_i 
-\int_\cB\iota_{\hat \xi}\iota_{\hat \varphi}\Sigma_i   \rd_A F^i { -\tfrac12 \int_S \iota_{\hat{\xi}}\iota_{\hat{\varphi}} F^i \wedge   (\e \times  \e)_i }\n\\
& \simeq& D_{[\hat\xi,\hat\varphi]}+G_{\iota_{\hat \xi}\iota_{\hat \varphi}F} \,.
\ea
In the last line we have used the Bianchi identity---or magnetic Gauss law---$\rd_{A}F=0$ and the definition \eqref{eqn:defGint} of the electric gauge generator in terms of the bulk electric Gauss law and boundary simplicity constraint, leading to the gauge generator $G_{\iota_{\hat \xi}\iota_{\hat \varphi}F}$ which generates gauge transformation with gauge parameter $ \iota_{\hat \xi}\iota_{\hat \varphi}F$.
This is consistent with the action of covariant diffeomorphism since the covariant Cartan calculus implies that 
\be
[L_{\hat{\xi}},L_{\hat{\varphi}}] = L_{[ \hat{\xi},\hat{\varphi}]} + \iota_{\hat \xi}\iota_{\hat \varphi}F\times\,,
\ee
showing how the connection curvature $F$ deforms the commutator of the covariant Lie derivatives.

\section{Conclusion}
We have revisited general relativity in its first order formulation in terms of frame field and connection in the presence of boundaries. 
We have seen that the gravity bulk phase space needs to be appropriately extended by a set of boundary edge modes to allow for arbitrary frame field variations on the boundary. 
We have shown that, quite remarkably, the kinematical constraints can be understood as conservation of boundary charges. We have demonstrated that these charges are the hamiltonian generators of electric gauge transformations and translations. We have proven that these charges form a closed Poincar\'e algebra, as anticipated in \cite{Freidel:2019ees}.

One has to wonder whether we can extend these results to the dynamical constraints. It has already been established in 
\cite{Hopfmuller:2018fni}, that Einstein equations projected along null surfaces can be understood, in the metric formulation, as conservation equations for charges aspects associated with energy and momenta. The canonical structure of these charges hasn't been revealed yet. AN relevant analysis in the first order as recently been performed in \cite{Wieland:2019hkz}

At this point, a fascinating question would be whether magnetic gauge transformations can also be included or not as phase space transformation. The physical implications of including the dual magnetic sector in the boundary phase space have been explored in \cite{Freidel:2018fsk} for electromagnetism. One may expect the magnetic charges to play an equally important role also in the case of gravity in order to unravel the full boundary symmetry group.
Indeed, although the Bianchi law $\rd_{A}F=0$ is a purely geometric identity, it is tempting to interpret it as a magnetic Gauss law. In this scenario,  the integrated  generator $F_{\beta}=\int \rd_{A}\beta^{i}\wedge F_{i}[A]$ would play the role of the generator of magnetic gauge transformations. One expects, following \cite{Freidel:2018fsk},  that it is necessary to introduce magnetic edge modes to render this generator differentiable.
It seems natural to expect these magnetic edge modes to be encoded in a boundary connection field.
 Either by providing with a Chern-Simons-like boundary symplectic term \cite{Ashtekar:1999wa, Engle:2010kt, Ghosh:2014rra} or with a  mixed term coupling the boundary connection to the boundary frame field  \cite{Pranzetti:2014tla} as was derived in the case of isolated horizon boundary conditions.
%

Similarly if one wants to ensure that the translations respect the boundary Gauss law, one would also have, according to the general philosophy of boundary charges and symmetries \cite{Donnelly:2016auv},  to allow for non-trivial  translational edge modes. This is left to future investigation.

\section*{Acknowledgements}
Research at Perimeter Institute for Theoretical Physics is supported in part by the
Government of Canada through NSERC and by the Province of Ontario through MRI.

 \bibliographystyle{bib-style}
 \bibliography{Gravity-Charge-final.bib}

\end{document}